\documentclass[conference]{IEEEtran}
\usepackage{amssymb,amsthm,amsmath,epsfig,latexsym,graphicx,bm,nccmath}
\usepackage{mathtools}

\usepackage{epstopdf}
\usepackage{cite}
\usepackage{amsfonts}
\usepackage[ruled, lined, linesnumbered, commentsnumbered, longend]{algorithm2e}
\usepackage{textcomp}
\usepackage{xcolor}
\def\BibTeX{{\rm B\kern-.05em{\sc i\kern-.025em b}\kern-.08em
		T\kern-.1667em\lower.7ex\hbox{E}\kern-.125emX}}
\usepackage{caption}	
\usepackage{subcaption}
\usepackage[utf8x]{inputenc}
\usepackage{etoolbox}
\usepackage{booktabs}
\usepackage[top=0.71in,bottom=1in,left=0.625in,right=0.625in]{geometry}

\usepackage[skip=0pt]{caption}
\usepackage[font=footnotesize,skip=0pt]{subcaption}
\usepackage{easyReview} 

\makeatletter
\patchcmd{\@maketitle}
{\addvspace{0.5\baselineskip}\egroup}
{\addvspace{-1\baselineskip}\egroup}
{}
{}
\makeatother

\begin{document}
	
	\title{Priority-Based Bandwidth Allocation in Network Slicing-Enabled Cell-Free Massive MIMO Systems}
	
	\author{\IEEEauthorblockN{Manobendu Sarker\IEEEauthorrefmark{1} and Soumaya Cherkaoui\IEEEauthorrefmark{2}} 
		\IEEEauthorblockA{Department of Computer and Software Engineering, Polytechnique Montr\'{e}al, Canada\\
			Email: \{manobendu.sarker\IEEEauthorrefmark{1}, soumaya.cherkaoui\IEEEauthorrefmark{2}\}@polymtl.ca}
	}
	
	\maketitle
	
	\begin{abstract}
		This paper addresses joint admission control and per-user equipment (UE) bandwidth allocation to maximize weighted sum-rate in network slicing-enabled user-centric cell-free (CF) massive multiple-input multiple-output (mMIMO) systems when aggregate quality-of-service (QoS) demand may exceed available bandwidth. Specifically, we optimize bandwidth allocation while satisfying heterogeneous QoS requirements across enhanced mobile broadband (eMBB) and ultra-reliable low-latency communication (URLLC) slices in the uplink. The formulated problem is NP-hard, rendering global optimality computationally intractable. We decompose it into two sub-problems and solve them via computationally efficient heuristics within a sequential framework. We propose (i) a hierarchical admission control scheme that selectively admits UEs under bandwidth scarcity, prioritizing URLLC to ensure latency-sensitive QoS compliance, and (ii) an iterative gradient-based bandwidth allocation scheme that transfers bandwidth across slices guided by marginal utility and reallocates resources within slices. Simulation results demonstrate that the proposed scheme achieves near-optimal performance, deviating from an interior point solver-based benchmark by at most 2.2\% in weighted sum-rate while reducing runtime by 99.7\%, thereby enabling practical real-time deployment. Compared to a baseline round-robin scheme without admission control, the proposed approach achieves up to 1085\% and 7\% higher success rates for eMBB and URLLC slices, respectively, by intentionally sacrificing sum-rate to guarantee QoS. Sensitivity analysis further reveals that the proposed solution adapts effectively to diverse eMBB/URLLC traffic compositions, maintaining 47--51\% eMBB and 93--94\% URLLC success rates across varying load scenarios, confirming its robustness for resource-constrained large-scale deployments.  
	\end{abstract}
	
	\vspace{-1mm}	
	\section{Introduction}	
	The next-generation wireless systems aim to support diverse applications with stringent quality-of-service (QoS) demands, primarily represented by enhanced mobile broadband (eMBB) and ultra-reliable low-latency communication (URLLC) services \cite{10190330}. These services exhibit fundamentally different characteristics: URLLC applications such as industrial automation and autonomous vehicles require strict latency bounds (e.g., 1 ms) and ultra-high reliability (99.999\%) \cite{3gpp38913}, where delays can cause critical failures, whereas eMBB traffic is elastic and delay-tolerant, allowing scheduling flexibility \cite{she2021tutorial}. Furthermore, URLLC packets arrive sporadically with stringent deadlines, while eMBB demands sustained throughput over longer timescales \cite{she2021tutorial}. Consequently, URLLC-first resource allocation ensures latency-critical traffic receives immediate access while eMBB user equipments (UEs) tolerate brief interruptions, enabling efficient coexistence under resource scarcity. Network slicing (NS) facilitates such service differentiation by partitioning physical infrastructure into logical slices optimized for distinct QoS objectives \cite{8320765}. To realize these benefits, cell-free (CF) massive multiple-input multiple-output (mMIMO) has emerged as a potential sixth generation (6G) paradigm \cite{Ngo2017}, where distributed access points (APs) cooperatively serve UEs, offering superior spectral efficiency (SE), uniform coverage, and enhanced reliability through user-centric (UC) connectivity \cite{Demir2021a}. However, combining NS with CF mMIMO introduces significant challenges in managing heterogeneous QoS requirements, particularly when prioritizing latency-sensitive URLLC traffic over eMBB services under limited spectrum.
	
	Existing works on bandwidth allocation in CF mMIMO systems primarily address feasible resource scenarios. For instance, \cite{9435371} allocates bandwidth using an interior-point method during the second stage of a bi-level resource allocation framework maximizing infrastructure revenue while preserving service quality, whereas \cite{10114604} allocates and adjusts bandwidth within slices using deep reinforcement learning to maximize cross-service utility. However, practical deployments often face infeasible scenarios where aggregated minimum bandwidth demand is more than available capacity. In such cases, admission control mechanisms that selectively reject low-priority or infeasible UEs become necessary to guarantee strict QoS for admitted users, particularly critical for URLLC applications where QoS violations can cause mission-critical failures. Nevertheless, admission control for QoS-constrained network slicing in CF mMIMO systems has not yet been explored in the literature.

    Building on this insight, we propose a joint admission control and bandwidth allocation framework with URLLC prioritization for NS-enabled uplink (UL) CF mMIMO systems under bandwidth scarcity where aggregate QoS demand may exceed available capacity. The framework operates in the presence of pilot contamination (PC) and ensures higher success rates for delay-sensitive URLLC UEs through selective admission and priority-aware resource allocation while maintaining system efficiency. To the best of our knowledge, joint priority-based admission control and bandwidth allocation in UC CF mMIMO systems have not yet been investigated in the literature. The main contributions are as follows: 
	
	\smallskip \noindent $\bullet$ We formulate a joint optimization problem for admission control and per-UE bandwidth allocation to maximize weighted sum-rate while satisfying heterogeneous QoS constraints across eMBB and URLLC slices in NS-enabled UL CF mMIMO systems under bandwidth scarcity where aggregate QoS demand may exceed available capacity; owing to the NP-hard nature of the problem, we propose a sequentially optimization approach that decomposes it into two sub-problems and solves them in cascade.
	
	\smallskip \noindent $\bullet$ We design two computationally efficient heuristics: a URLLC-prioritized hierarchical admission control that selectively admits UEs under resource constraints, and an iterative marginal utility-guided bandwidth allocation scheme that jointly ensure QoS compliance for latency-sensitive UEs while maintaining system efficiency.
	
	\smallskip \noindent $\bullet$ Numerical analysis confirms that the proposed solution achieves near-optimal performance and robust QoS satisfaction across diverse traffic compositions with significantly reduced computational complexity compared to an interior-point solver-based solution, demonstrating practical viability for bandwidth-constrained large-scale deployments.
	
	\vspace{-2mm}
	\section{System Model}
	Our considered CF mMIMO network comprises $M$ distributed APs, each equipped with $N$ antennas, jointly serving $K$ geographically dispersed single-antenna UEs. A central processing unit (CPU) orchestrates cooperative transmission via fronthaul links connecting all APs. UEs are partitioned into $S=2$ network slices corresponding to eMBB and URLLC services. We denote by $\mathcal{M}=\{1,\dots,M\}$, $\mathcal{K}=\{1,\dots,K\}$, and $\mathcal{S}=\{1,\dots,S\}$ the index sets for APs, UEs, and slices, respectively. For each slice $s\in\{\mathrm{eMBB},\mathrm{URLLC}\}$, the subset $\mathcal{K}_s\subseteq\mathcal{K}$ contains the UEs belonging to that slice. While each UE $k$ is associated with a single slice $s_k\in\mathcal{S}$, it may be cooperatively served by multiple APs through the subset $\mathcal{V}_k \subseteq \mathcal{M}$. In contrast to the service-type-restricted AP assignment in \cite{10198770}, our architecture permits each AP to simultaneously serve both eMBB and URLLC UEs, thereby enhancing spatial diversity and load balancing. The total system bandwidth $B$ is partitioned across slices such that $\sum_s B_s\le B$, where $B_s$ denotes the bandwidth budget for slice $s$, and each UE $k$ receives an individual allocation $b_k\ge0$.  Priority weights $w_k>0$ assigned to each UE $k$ differentiate QoS requirements across heterogeneous services.
	\vspace{-2mm}
	\subsection{Channel Propagation and Estimation}
	
	The wireless link between UE $k \in \mathcal{K}$ and AP $m \in \mathcal{M}$ is characterized by $\mathbf{g}_{k,m} = \sqrt{\beta_{k,m}} \, \mathbf{h}_{k,m}$, where $\mathbf{h}_{k,m} \in \mathbb{C}^{N \times 1}$ denotes the small-scale fading with distribution $\mathcal{CN}(\mathbf{0}, \mathbf{I}_N)$, and $\beta_{k,m}$ represents the large-scale fading coefficient (LSFC). The LSFC incorporates path loss and shadowing via $\beta_{k,m} = 10^{\frac{\text{PL}_{k,m}}{10}} \times 10^{\frac{\sigma_{\mathrm{sh}} z_{k,m}}{10}}$, where $\text{PL}_{k,m}$ (dB) follows the three-slope model of \cite{Ngo2017}, $\sigma_{\mathrm{sh}}$ (dB) is the shadow fading standard deviation, and $z_{k,m} \sim \mathcal{N}(0,1)$ models log-normal shadowing. We assume perfect LSFC knowledge at the CPU and all APs.
	
	Each coherence interval of $\tau_c$ symbols allocates $\tau_p$ symbols for pilot transmission and the remainder for data. All UEs simultaneously transmit pilot sequences $\boldsymbol{\psi}_k \in \mathbb{C}^{\tau_p \times 1}$ with $\|\boldsymbol{\psi}_k\|^2 = 1$ assigned from a $\tau_p$ orthogonal sequences. Since $K \gg \tau_p$ in realistic deployments, pilot reuse introduces PC. We adopt minimum mean square error (MMSE) estimation to obtain channel estimates $\hat{\mathbf{g}}_{k,m}$, whose quality is quantified by $\gamma_{k,m} \triangleq \mathbb{E} \left[ \| \hat{\mathbf{g}}_{k,m} \|^2 \right] = \sqrt{\tau_p \rho_p \, \eta_k^p} \, \beta_{k,m} \, c_{k,m}$, where $\rho_p$ is the normalized pilot signal-to-noise ratio (SNR), $\eta_k^p \in (0,1]$ is UE $k$'s pilot power control coefficient \cite{Ngo2017}, and
	\begin{equation}
		c_{k,m} \triangleq \frac{\sqrt{\tau_p \rho_p} \, \beta_{k,m} \sqrt{\eta_k^p}}{\tau_p \rho_p \sum\limits_{j \in \mathcal{K}} \beta_{j,m} \eta_j^p |\boldsymbol{\psi}_k^H \boldsymbol{\psi}_j|^2 + 1}.
		\label{MMSE}
	\end{equation}
	
	\subsection{Uplink Data Transmission and Latency Characterization}
	\label{Uplink Data Transmission and Latency Characterization}
	
	Each UE $k \in \mathcal{K}$ transmits data symbols to APs in $\mathcal{V}_k$\footnote{Every AP captures signals from all UEs, yet it only processes the transmissions of its designated UEs as determined by the UE–AP association policy.} during UL data transmission which forwards received signals to the CPU for centralized joint decoding using $\hat{\mathbf{g}}_{k,m}$. For eMBB services, Shannon capacity provides an accurate rate model, whereas URLLC short packets require finite blocklength analysis \cite{10114604}. The achievable rate for both service classes is presented in \eqref{rate_eq} at the top of the next page, where $\theta$ is target decoding error probability, ${Q}^{-1}(\cdot)$ is the inverse Gaussian Q-function, and $V_k = 1-(1+\text{SINR}_k)^{-2}$ captures channel dispersion. Following \cite{1193803}, the signal-to-interference-plus-noise ratio (SINR) for UE $k$ is shown in \eqref{SINR_closed}, where $\rho_d$ is normalized UL data SNR and $\eta_{k}^d \in (0,1]$ is the data power control coefficient.
	
	\begin{figure*}[tb]
		\begin{align}
			R_{k}= b_{k} \text{SE}_k =&\left\lbrace \begin{array}{llll} b_{k} \left(1 - \frac{\tau_p}{\tau_c} \right) \log_2 \left(1 + \text{SINR}_{k} \right), & \forall k \in \mathcal{K}_{\mathrm{eMBB}}, \\ b_{k} \left(1 - \frac{\tau_p}{\tau_c} \right) \left[\log_2 \left(1 + \text{SINR}_{k} \right)-\sqrt{\frac{V_k}{L_k}} \frac{{{Q}^{-1}}\left(\theta \right)}{\ln 2} \right], & \forall k \in \mathcal{K}_{\mathrm{URLLC}}. \end{array} \right.
			\label{rate_eq}
		\end{align}
		\vspace{-6pt}
		\noindent\rule{\textwidth}{1pt}
	\end{figure*}
	
	\begin{figure*}[tb]
		\begin{align}
			\text{SINR}_k = \frac{N^2\rho_d\eta_k^d\left(\sum_{m \in \mathcal{V}_k} \ \gamma_{k,m}\right)^2}{N\rho_d\sum_{k'=1}^{K}\eta_{k'}^d\sum_{m \in \mathcal{V}_k} \ \gamma_{k,m}\beta_{k',m}+N^2\rho_d \sum_{k'\neq k}^{K}\eta_{k'}^d|\psi_k^H\psi_{k'}|^2\left(\sum_{m \in \mathcal{V}_k} \ \gamma_{k,m} \sqrt{\frac{\eta_{k'}^p}{\eta_k^p}} \frac{\beta_{k',m}}{\beta_{k,m}}\right)^2+N\sum_{m \in \mathcal{V}_k} \ \gamma_{k,m}}.
			\label{SINR_closed}
		\end{align}
		\vspace{-6pt}
		\noindent\rule{\textwidth}{1pt}
	\end{figure*}
	
	To model URLLC end-to-end latency, we employ an $M/M/1$ queueing framework \cite{10772596}. For UE $k \in \mathcal{K}_{\mathrm{URLLC}}$, packets arrive at rate $\lambda_{k}$ (packets/s) and are served at rate $\mu_{k} = R_{k} / L_{k}$ (packet size in bytes). Under stability condition $\mu_{k} > \lambda_{k}$ \cite{kleinrock1975queueing}, the total delay is $D_{k} = \frac{1}{\mu_{k} - \lambda_{k}} = \frac{1}{\frac{R_{k}}{L_{k}} - \lambda_{k}}$, combining queueing and transmission delays. To satisfy URLLC latency requirements, we enforce $D_{k} \leq D_{k}^{\mathrm{max}}$, which translates to the minimum rate condition $\frac{R_{k}}{L_{k}} \geq \lambda_{k} + \frac{1}{D_{k}^{\mathrm{max}}}, \forall k \in \mathcal{K}_{\mathrm{URLLC}}$, ensuring delays remain within bounds for mission-critical service delivery.
	\vspace{-1mm}	
	\section{Problem Formulation}
	
	We formulate the joint admission and bandwidth allocation problem to maximize weighted sum-rate under fixed bandwidth $B$ and QoS constraints. The optimization problem is formulated as follows:
	\begin{subequations}
		\small
		\label{eq:master_problem}
		\begin{align}
			\mathcal{P}_0: \ \underset{\substack{\{\alpha_k\} \in \{0,1\},\\ \{b_k\}, \{B_s\} \geq 0}}{\text{max}} \quad & \sum_{k=1}^{K} \alpha_k w_k b_k \text{SE}_k \label{eq:master_obj} \\
			\text{s.t.} \quad & \sum_{k \in \mathcal{K}_s} b_k \leq B_s, \quad \forall s \in \mathcal{S}, \label{eq:slice_capacity} \\
			& \sum_{s=1}^{S} B_s = B, \label{eq:total_bw} \\
			& b_k \geq \alpha_k b_{k}^{\min}, \quad \forall k, \label{eq:min_bw} \\
			& B_s \geq \sum_{k \in \mathcal{K}_s} \alpha_k b_{k}^{\min}, \quad \forall s \in \mathcal{S}, \label{eq:slice_min}
		\end{align}
	\end{subequations}
	where $\alpha_k \in \{0,1\}$ indicates admission and $b_k^{\min} = R_k^{\min}/\text{SE}_k$ is the minimum bandwidth for UE $k$ to meet QoS. Note that the QoS constraint for eMBB UEs is minimum rate $R_k^{\min}$, whereas for URLLC UEs it is maximum delay $D_{k}^{\mathrm{max}}$.
	
	Constraints~\eqref{eq:slice_capacity} and~\eqref{eq:total_bw} enforce per-slice capacity limits and total bandwidth partitioning (equality prevents waste), respectively. Constraint~\eqref{eq:min_bw} couples admission to provisioning: admitted UEs ($\alpha_k = 1$) receive bandwidth $\geq b_k^{\min}$ and rejected UEs ($\alpha_k = 0$) get no bandwidth allocation. Constraint~\eqref{eq:slice_min} ensures that each slice reserves capacity for its admitted UEs. While objective~\eqref{eq:master_obj} prioritizes URLLC via $w_k^{\text{URLLC}} \gg w_k^{\text{eMBB}}$, weights alone cannot guarantee URLLC admission before eMBB under scarcity. 
	
	Problem $\mathcal{P}_0$ is a mixed-integer bilinear program, where binary $\boldsymbol{\alpha}$ variable induces combinatorial complexity of $2^K$, and product terms $\alpha_k b_k$ render~\eqref{eq:master_obj} and~\eqref{eq:min_bw} non-convex. However, encoding strict lexicographic priority via constraints would require $\mathcal{O}(|\mathcal{K}_{\text{URLLC}}| \times |\mathcal{K}_{\text{eMBB}}|)$ binary conditions, further exacerbating complexity.
	\vspace{-2mm}
	\section{Proposed Solution}
	\label{sec:proposed}
	
	Problem $\mathcal{P}_0$ exhibits bilinear coupling between admission decisions $\{\alpha_k\}$ and bandwidth allocations $\{b_k\}$ in~\eqref{eq:master_obj}, \eqref{eq:min_bw}, and~\eqref{eq:slice_min}, rendering direct solution intractable for large systems. For computational efficiency, we decompose $\mathcal{P}_0$ into two tractable sub-problems and solve sequentially.
	
	\subsection{Sub-problem 1: Hierarchical Admission Control}
	For given slice bandwidths $\{B_s\}$, we determine the admitted UE set $\mathcal{K}_{\text{adm}} \subseteq \mathcal{K}$ via a priority-ordered multi-dimensional knapsack:
	\begin{subequations}
		\small
		\label{eq:admission_problem}
		\begin{align}
			\mathcal{P}_1: \underset{\boldsymbol{\alpha} \in \{0,1\}}{\text{max}}  & \quad \sum_{k=1}^{K} \alpha_k \Gamma_k \label{eq:adm_obj} \\
			\text{s.t.} \quad & \eqref{eq:slice_min},  \\
			& \alpha_k \succ \alpha_j, \ \forall k \in \mathcal{K}_{\text{URLLC}}, \forall j \in \mathcal{K}_{\text{eMBB}}, \label{eq:priority}
		\end{align}
	\end{subequations}
	where $\Gamma_k = w_k \cdot \text{SE}_k$ is the efficiency metric (weighted SE), and constraint~\eqref{eq:priority} encodes hierarchical priority: all feasible URLLC UEs are admitted before any eMBB UE, regardless of $\Gamma_k$. Problem $\mathcal{P}_1$ is NP-hard~\cite{Kellerer2004Knapsack} and explicit encoding of~\eqref{eq:priority} requires $\mathcal{O}(|\mathcal{K}_{\text{URLLC}}| \times |\mathcal{K}_{\text{eMBB}}|)$ binary conditions. To solve $\mathcal{P}_1$ efficiently, we propose a two-stage admission control algorithm detailed below.
	
	\subsubsection{Proposed Hierarchical Priority Admission Control}
	\label{sec:admission}
	
	Our hierarchical admission control determines $\mathcal{K}_{\text{adm}}$ with URLLC-first priority, ensuring sufficient bandwidth for latency-sensitive QoS even at eMBB's expense. The admission process consists of two stages, outlined in Algorithm~\ref{alg:admission}.
	
	\smallskip \noindent $\bullet$ \textbf{Stage~1} (Lines~\ref{line:urllc_start}--\ref{line:urllc_end}) allocates URLLC UEs access to $B_{\text{URLLC}}^{\max} = B - B_{\text{eMBB}}^{\min}$, where $B_{\text{eMBB}}^{\min}$ is a configurable eMBB soft floor. Only URLLC candidates with $b_k^{\min} \leq B_{\text{URLLC}}^{\max}$ are considered for admission (Line~\ref{line:urllc_filter}). These feasible candidates are ranked by their efficiency metric $\Gamma_k = w_k \cdot \text{SE}_k$ (Line~\ref{line:urllc_sort}), balancing priority and channel quality. Finally, the greedy knapsack iteration admits UE $k$ if cumulative demand stays within $B_{\text{URLLC}}^{\max}$ (Lines~\ref{line:urllc_knapsack_start}--\ref{line:urllc_knapsack_end}), yielding the URLLC admitted set $\mathcal{K}_{\text{URLLC}}^{\text{adm}}$ and bandwidth consumption $B_{\text{URLLC}}^{\text{used}}$ upon completion.
	
	\smallskip \noindent $\bullet$ \textbf{Stage~2} (Lines~\ref{line:embb_start}--\ref{line:embb_end}) allocates residual bandwidth $B_{\text{residual}} = B - B_{\text{URLLC}}^{\text{used}}$ to eMBB UEs. eMBB candidates with $b_k^{\min} \leq B_{\text{residual}}$ are filtered in Line~\ref{line:embb_filter} and then sorted by $\Gamma_k$ in Line~\ref{line:embb_sort}. After that, the algorithm selects eMBB UEs as in \textbf{Stage~1} until exhausting $B_{\text{residual}}$ (Lines~\ref{line:embb_knapsack_start}--\ref{line:embb_knapsack_end}), yielding $\mathcal{K}_{\text{eMBB}}^{\text{adm}}$ and $B_{\text{eMBB}}^{\text{used}}$.
	
	\subsubsection{Computational Complexity}
	
	Algorithm~\ref{alg:admission} is dominated by sorting in both stages. \textbf{Stage~1} incurs $\mathcal{O}(|\mathcal{K}_{\text{URLLC}}|)$ for filtering (Line~\ref{line:urllc_filter}) and $\mathcal{O}(|\mathcal{F}_{\text{URLLC}}| \log |\mathcal{F}_{\text{URLLC}}|)$ for sorting (Line~\ref{line:urllc_sort}), where $|\mathcal{F}_{\text{URLLC}}| \leq K$. The selection process (Lines~\ref{line:urllc_knapsack_start}--\ref{line:urllc_knapsack_end}) requires $\mathcal{O}(K)$ with constant-time per-UE decisions. \textbf{Stage~2} performs identical operations on eMBB candidates, contributing $\mathcal{O}(K \log K)$. Since sorting dominates filtering and greedy selection, the overall complexity is $\mathcal{O}(K \log K)$.
	
	\begin{algorithm}[t]
		\caption{Proposed Hierarchical Priority Admission Control}
		\label{alg:admission}
		\small
		\DontPrintSemicolon
		\KwIn{$\{\text{SE}_k, w_k, b_{k}^{\min}, \Gamma_k\}$,  $\mathcal{K}_{\text{URLLC}}, \mathcal{K}_{\text{eMBB}}$,  $B$,  $B_{\text{eMBB}}^{\min} = 0.2B$}
		\KwOut{$\mathcal{K}_{\text{adm}}$, $\mathbf{B}_s^{\min} = [B_{\text{URLLC}}^{\text{used}}, B_{\text{eMBB}}^{\text{used}}]^\top$}
		
		\label{line:urllc_start}
		$B_{\text{URLLC}}^{\max} \gets B - B_{\text{eMBB}}^{\min}$ \;
		$\mathcal{F}_{\text{URLLC}} \gets \{k \in \mathcal{K}_{\text{URLLC}} : b_{k}^{\min} \leq B_{\text{URLLC}}^{\max} \land b_{k}^{\min} < \infty\}$ \label{line:urllc_filter} \;
		Sort $\mathcal{F}_{\text{URLLC}}$ in descending order of $\Gamma_k$ $\rightarrow$ $\mathcal{F}_{\text{URLLC}}^{\text{sorted}}$ \label{line:urllc_sort} \;
		$B_{\text{cum}} \gets 0$, \quad $\mathcal{K}_{\text{URLLC}}^{\text{adm}} \gets \emptyset$ \;
		\label{line:urllc_knapsack_start}
		\ForEach{$k \in \mathcal{F}_{\text{URLLC}}^{\text{sorted}}$}{
			\If{$B_{\text{cum}} + b_{k}^{\min} \leq B_{\text{URLLC}}^{\max}$}{
				$\mathcal{K}_{\text{URLLC}}^{\text{adm}} \gets \mathcal{K}_{\text{URLLC}}^{\text{adm}} \cup \{k\}$ , \ $B_{\text{cum}} \gets B_{\text{cum}} + b_{k}^{\min}$ \;
			}
		}
		\label{line:urllc_knapsack_end}
		$B_{\text{URLLC}}^{\text{used}} \gets B_{\text{cum}}$ \;
		\label{line:urllc_end}
		
		\BlankLine
		$B_{\text{residual}} \gets B - B_{\text{URLLC}}^{\text{used}}$ \label{line:embb_start} \;
		\label{line:violation_end}
		$\mathcal{F}_{\text{eMBB}} \gets \{k \in \mathcal{K}_{\text{eMBB}} : b_{k}^{\min} \leq B_{\text{residual}} \land b_{k}^{\min} < \infty\}$ \label{line:embb_filter} \;
		Sort $\mathcal{F}_{\text{eMBB}}$ in descending order of $\Gamma_k$ $\rightarrow$ $\mathcal{F}_{\text{eMBB}}^{\text{sorted}}$ \label{line:embb_sort} \;
		$B_{\text{cum}} \gets B_{\text{URLLC}}^{\text{used}}$, \quad $\mathcal{K}_{\text{eMBB}}^{\text{adm}} \gets \emptyset$ \;
		\label{line:embb_knapsack_start}
		\ForEach{$k \in \mathcal{F}_{\text{eMBB}}^{\text{sorted}}$}{
			\If{$B_{\text{cum}} + b_{k}^{\min} \leq B$}{
				$\mathcal{K}_{\text{eMBB}}^{\text{adm}} \gets \mathcal{K}_{\text{eMBB}}^{\text{adm}} \cup \{k\}$, \				$B_{\text{cum}} \gets B_{\text{cum}} + b_{k}^{\min}$ \;
			}
		}
		\label{line:embb_knapsack_end}
		$B_{\text{eMBB}}^{\text{used}} \gets B_{\text{cum}} - B_{\text{URLLC}}^{\text{used}}$ 
		\label{line:embb_end},	$\mathcal{K}_{\text{adm}} \gets \mathcal{K}_{\text{URLLC}}^{\text{adm}} \cup \mathcal{K}_{\text{eMBB}}^{\text{adm}}$ \;
	\end{algorithm}
	
	\subsection{Sub-problem 2: Bandwidth Allocation}
	Given the admitted set $\mathcal{K}_{\text{adm}}$ from $\mathcal{P}_1$, the optimization problem of jointly optimizing slice bandwidths $\{B_s\}$ and per-UE allocations $\{b_k\}$ is formulated as:
	\begin{subequations}
		\small
		\label{eq:bandwidth_problem}
		\begin{align}
			\mathcal{P}_2:  \underset{\{{b}_k\}, \{{B}_s\} \geq 0}{\text{maximize}} \quad & \sum_{k \in \mathcal{K}_{\text{adm}}} w_k b_k \text{SE}_k \label{eq:bw_obj} \\
			\text{subject to} \quad & \sum_{k \in \mathcal{K}_s \cap \mathcal{K}_{\text{adm}}} b_k = B_s, \quad \forall s \in \mathcal{S}, \label{eq:bw_slice_eq} \\
			& \sum_{s=1}^{S} B_s = B, \label{eq:bw_total} \\
			& b_k \geq b_{k}^{\min}, \quad \forall k \in \mathcal{K}_{\text{adm}}, \label{eq:bw_min} \\
			& B_s \geq \sum_{k \in \mathcal{K}_s \cap \mathcal{K}_{\text{adm}}} b_{k}^{\min}, \quad \forall s \in \mathcal{S}. \label{eq:bw_slice_min} 
		\end{align}
	\end{subequations}
	
	While $\mathcal{P}_2$ is convex when $\{B_s\}$ is fixed, joint optimization of inter-slice and intra-slice allocations is challenging due to coupling constraint~\eqref{eq:bw_slice_eq}. We propose an iterative gradient-based approach (Algorithm~\ref{alg:bandwidth}) that alternates between bandwidth transfer across slices guided by marginal utility.
	\vspace{-1mm}
	\subsubsection{Proposed Bandwidth Allocation Scheme}
	
	Algorithm~\ref{alg:bandwidth} initializes by allocating each slice its minimum requirement $B_s^{\min}$ from Algorithm~\ref{alg:admission}, distributing the surplus $B_{\text{surplus}} = B - \sum_{s} B_s^{\min}$ proportionally to admitted UEs: $B_s = B_s^{\min} + B_{\text{surplus}} N_s / \sum_{s'} N_{s'}$, where $N_s = |\mathcal{K}_s \cap \mathcal{K}_{\text{adm}}|$ (Lines~\ref{line:init_start}--\ref{line:init_end}). Initial per-UE allocation (Line~\ref{line:init_alloc}) satisfies~\eqref{eq:bw_slice_eq}, \eqref{eq:bw_min}, and~\eqref{eq:bw_slice_min}, which is assigned by a quadratic allocation (QA) subroutine.
	
	The iterative loop (Lines~\ref{line:loop_start}--\ref{line:loop_end}) finalizes the bandwidth allocation. Algorithm~\ref{alg:bandwidth} uses another subroutine for calculating marginal utility, $\mu_s, \ s \in \mathcal{S}$ in Line~\ref{line:mu_calc}. From these $\mu_s$ values, the algorithm identifies donor (minimum $\mu_s$) and receiver (maximum $\mu_s$) slices (Line~\ref{line:donor}). The algorithm terminates if marginal utilities are balanced ($\mu_{s_{\text{rcv}}} \leq 1.05 \mu_{s_{\text{donor}}}$) in Line~\ref{line:conv_mu}. Otherwise, bandwidth $\Delta B$ is transferred from donor to receiver slice (Line~\ref{line:delta_b}), where $\Delta B$ is the average per-UE bandwidth in the donor slice, capped by the excess above its minimum requirement. Termination also occurs if $\Delta B $ is below a small threshold $\epsilon_{\text{transfer}}$ (Line~\ref{line:conv_delta}). After updating slice allocations (Line~\ref{line:transfer}), per-UE bandwidth is again recomputed with QA subroutine in Line~\ref{line:realloc}, and Line~\ref{line:eval} evaluates the new objective $\tilde{f} = \sum_{k \in \mathcal{K}_{\text{adm}}} w_k \tilde{b}_k \text{SE}_k$. The transfer is accepted if $\tilde{f}$ exceeds the previous objective (Lines~\ref{line:accept_start}--\ref{line:accept_end}); otherwise, it is rejected (Line~\ref{line:reject}). However, early stopping can also occur after $T_{\text{patience}}$ consecutive rejections (Line~\ref{line:patience}).  Algorithm~\ref{alg:bandwidth} employs two subroutines for per‑UE bandwidth calculation in Lines~\ref{line:init_alloc} and~\ref{line:realloc}, and marginal‑utility computation in Line~\ref{line:mu_calc}, described next.
	
	\textbf{Quadratic Allocation (QA) Subroutine:} The QA subroutine distributes bandwidth within each slice to satisfy the coupling constraint~\eqref{eq:bw_slice_eq} while respecting minimum requirements~\eqref{eq:bw_min} by minimizing deviation from weighted proportional targets:
	\begin{subequations}
		\small
		\label{eq:quadratic_allocation}
		\begin{align}
			\min_{\{b_k\} \geq 0} \quad & \sum_{k \in \mathcal{K}_s \cap \mathcal{K}_{\text{adm}}} \left(b_k - \bar{b}_k\right)^2 \\
			\text{s.t.} \quad & \eqref{eq:bw_slice_eq}, \eqref{eq:bw_min},
		\end{align}
	\end{subequations}
	where $\bar{b}_k = w_k B_s / \sum_{j \in \mathcal{K}_s \cap \mathcal{K}_{\text{adm}}} w_j$. This is implemented by assigning each UE $b_k = b_k^{\min}$, then distributing surplus $B_s - \sum_{k \in \mathcal{K}_s^{\text{adm}}} b_k^{\min}$ proportionally to $\phi_k = (w_k \text{SE}_k)^2$, requiring $\mathcal{O}(N_s)$ per slice or $\mathcal{O}(K)$ total. 
	
	\textbf{Marginal Utility (MU) Subroutine:}     
	This subroutine computes marginal utility for each slice at iteration $t$, defined as:
	\begin{equation}
		\label{eq:marginal_utility_def}
		\mu_s^{(t)} = \frac{\partial}{\partial B_s} \sum_{k \in \mathcal{K}_{\text{adm}}} w_k b_k^{(t)} \text{SE}_k \bigg|_{B_s = B_s^{(t)}}, \ s \in \mathcal{S},
	\end{equation}
	representing global objective improvement per unit bandwidth increase to slice $s$.
	The marginal utility~\eqref{eq:marginal_utility_def} is computed via finite-difference gradient approximation: it evaluates the objective under current allocations and under a perturbed bandwidth $\tilde{B}_s = B_s + \delta_B$ (with $\delta_B = 0.1$ MHz), computing $\mu_s = (f(\tilde{B}_s) - f(B_s))/\delta_B$. This requires two allocation calls at $\mathcal{O}(K)$ each, yielding $\mathcal{O}(K)$ per slice or $\mathcal{O}(SK)$ for all slices.
	
	\paragraph{Computational Complexity}
	Algorithm~\ref{alg:bandwidth} initialization requires $\mathcal{O}(S + K)$ for surplus distribution and initial allocation. Each iteration incurs $\mathcal{O}(SK)$ for marginal utility computation (Line~\ref{line:mu_calc}), $\mathcal{O}(S)$ for donor-receiver identification, and $\mathcal{O}(K)$ for reallocation and evaluation (Lines~\ref{line:init_alloc} and~\ref{line:realloc}), yielding a per-iteration complexity $\mathcal{O}(SK)$. Over at most $T_{\max}$ iterations, the overall computational complexity of Algorithm~\ref{alg:bandwidth} is $\mathcal{O}(T_{\max} SK)$.
	
	\begin{algorithm}[t]
		\caption{Proposed Bandwidth Allocation Algorithm}
		\label{alg:bandwidth}
		\small
		\DontPrintSemicolon
		
		\KwIn{$\mathcal{K}_{\text{adm}}$, $\mathbf{B}_s^{\min} = [B_1^{\min}, \ldots, B_S^{\min}]^\top$, $\{w_k, \text{SE}_k, b_k^{\min}\}_{k \in \mathcal{K}_{\text{adm}}}$, $B$, $T_{\max}$, $T_{\text{patience}}$, $\epsilon_{\text{transfer}}$}
		\KwOut{$\{B_s\}^*$, $\{b_k\}^*$}
		
		$B_s \gets B_s^{\min}, \; \forall s \in \mathcal{S}$ \label{line:init_start}, $B_{\text{surplus}} \gets B - \sum_{s=1}^S B_s^{\min}$ \label{line:surplus}\;
		\label{line:dist_start}
		
		$N_s \gets |\mathcal{K}_s \cap \mathcal{K}_{\text{adm}}|$, 
		$B_s \gets B_s + B_{\text{surplus}} \cdot \frac{N_s}{\sum_{s'=1}^S N_{s'}}, \ \forall s \in \mathcal{S}$ \;
		
		\label{line:dist_end}
		Calculate $\{b_k\}$ using QA Subroutine \label{line:init_alloc}\;
		$f^{(0)} \gets \sum_{k \in \mathcal{K}_{\text{adm}}} w_k b_k \text{SE}_k$ \;
		$(\{B_s\}^*, \{b_k\}^*, f^*) \gets (\{B_s\}, \{b_k\}, f^{(0)})$ \;
		$t \gets 0$, $t_{\text{no\_impr}} \gets 0$ \;
		\label{line:init_end}
		
		\BlankLine
		\While{$t < T_{\max}$}{\label{line:loop_start}
			$t \gets t + 1$ \;
			Calculate $\{\mu_s\}$ using MU Subroutine \label{line:mu_calc}\;
			$s_{\text{donor}} \gets \arg\min_{s \in \mathcal{S}} \mu_s$ \label{line:donor}, 		$s_{\text{rcv}} \gets \arg\max_{s \in \mathcal{S}} \mu_s$ \label{line:rcv}\;
			\If{$\mu_{s_{\text{rcv}}} \leq 1.05 \cdot \mu_{s_{\text{donor}}}$}{\label{line:conv_mu}
				\textbf{break}\;
			}
			$\bar{b}_{\text{donor}} \gets B_{s_{\text{donor}}} / N_{s_{\text{donor}}}$ \;
			$\Delta B \gets \min\left(\bar{b}_{\text{donor}}, B_{s_{\text{donor}}} - B_{s_{\text{donor}}}^{\min}\right)$ \label{line:delta_b}\;
			\If{$\Delta B < \epsilon_{\text{transfer}}$}{\label{line:conv_delta}
				\textbf{break}\;
			}
			$\tilde{B}_{s_{\text{donor}}} \gets B_{s_{\text{donor}}} - \Delta B$, 
			$\tilde{B}_{s_{\text{rcv}}} \gets B_{s_{\text{rcv}}} + \Delta B$ \label{line:transfer}\;
			$\tilde{B}_{s'} \gets B_{s'}, \; \forall s' \notin \{s_{\text{donor}}, s_{\text{rcv}}\}$ \;
			Calculate $\{\tilde{b}_k\}$ using QA Subroutine with $\{\tilde{B}_s\}$ \label{line:realloc}\;
			$\tilde{f} \gets \sum_{k \in \mathcal{K}_{\text{adm}}} w_k \tilde{b}_k \text{SE}_k$ \label{line:eval}\;
			\eIf{$\tilde{f} > f^{(t-1)}$}{\label{line:accept_start}
				$\{B_s\} \gets \{\tilde{B}_s\}$, $\{b_k\} \gets \{\tilde{b}_k\}$,  $f^{(t)} \gets \tilde{f}$ \;
				\If{$\tilde{f} > f^*$}{
					$\{B_s\}^* \gets \{\tilde{B}_s\}$, $\{b_k\}^* \gets \{\tilde{b}_k\}$, $f^* \gets \tilde{f}$ \;
				}
				$t_{\text{no\_impr}} \gets 0$ \;\label{line:accept_end}
			}{
				$\{B_s\} \gets \{B_s\}$, $\{b_k\} \gets \{b_k\}$,  $f^{(t)} \gets f^{(t-1)}$ \;
				$t_{\text{no\_impr}} \gets t_{\text{no\_impr}} + 1$ \label{line:reject}\;
			}
			\If{$t_{\text{no\_impr}} \geq T_{\text{patience}}$}{\label{line:patience}
				\textbf{break}\;
			}
		}\label{line:loop_end}
	\end{algorithm}
	
	\vspace{-2mm}
	\section{Numerical Analysis}
	\label{Numerical Analysis}
	
	We evaluate our proposed schemes in a NS-enabled CF mMIMO network over $1 \times 1$ km$^2$ area, where $M=100$ APs equipped with $N = 4$ antennas serve $K$ single-antenna UEs. Spatial distribution of both APs and UEs follows a uniform model, and the wrap‑around technique \cite{Bjornson2017} is applied to remove boundary effects. The system operates with coherence parameters $\tau_p = 10$ and $\tau_c= 200$, while pilot and data transmit powers are set to $\rho_p=\rho_d= 100~\text{mW/Np}$ \cite{Ngo2017}. Log-normal shadow fading with standard deviation $\sigma_{sh}= 8$ dB is adopted. To intensify PC, pilots are randomly distributed and their power is assigned per \cite{Sark2301:Pilot}, whereas UL data is transmitted with open-loop power control. The transmit power for both pilot and data signals is limited to 100 mW. We adopt the UE-AP association strategy from~\cite{sarker2023access}.
	
	The total bandwidth is configured to $B = 80$ MHz to accommodate eMBB and URLLC QoS requirements. Unless otherwise specified, the UE mix comprises 70$\%$ eMBB and 30$\%$ URLLC UEs. URLLC traffic is modeled with dynamic packet arrivals characterized by packet sizes $L_k \in [32,64]$ bytes, arrival rates $\lambda_k \in [5,25]$ packets/s, delay budgets $D_k^{\max} \in [1,5]$ ms, and priority weights $w_k \in [2,4]$, under a decoding error constraint $\theta = 10^{-5}$. The eMBB population consists of 30$\%$ premium UEs ($R_k^{\min} \in [5,10]$ Mbps, $w_k=1.5$) and 70$\%$ standard UEs ($R_k^{\min} \in [1,3]$ Mbps, $w_k=1.0$). For Algorithm~\ref{alg:bandwidth}, we set $T_{\max} = 50$ iterations, $T_{\text{patience}} = 5$, and $\epsilon_{\text{transfer}} = 0.01$ MHz. 
	
	\subsubsection{Average Weighted Sum-Rate Performance}
	\label{Average Weighted Sum-rate Performance}
	
\begin{figure}[tb]	
		\centering
		\includegraphics[scale=0.2]{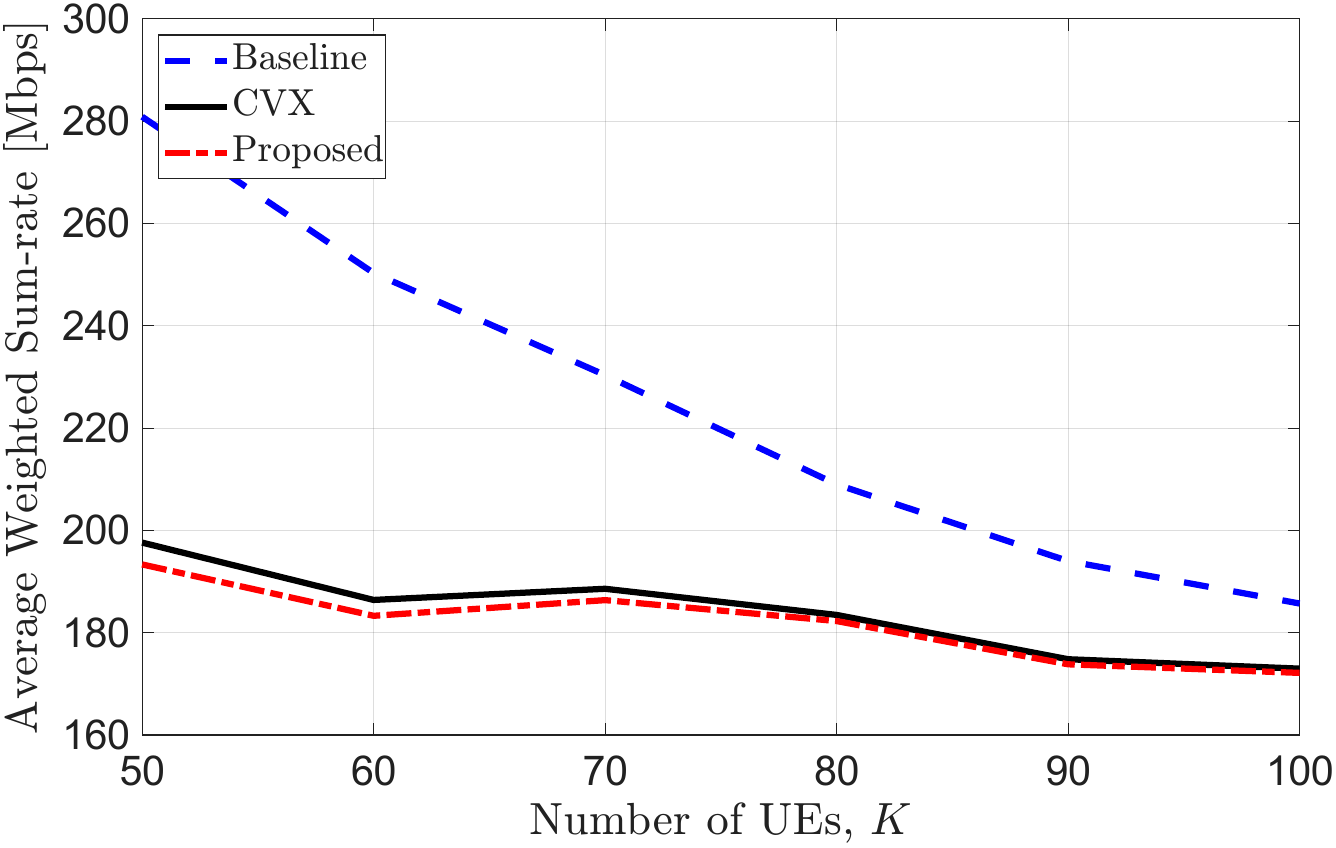}
		\caption{Comparison of average weighted sum-rate achieved by different schemes for varying UEs $K$ with  $M = 100$ APs.}
		\label{sumrate}
	\end{figure}

	Fig.~\ref{sumrate} compares the average weighted sum-rate performance among three schemes: the proposed scheme (`Proposed'), a `CVX' scheme employing CVX, a disciplined convex programming solver-based bandwidth allocation, and a `Baseline' scheme using round-robin bandwidth allocation without admission control. Both the CVX and Proposed schemes utilize the proposed priority-based admission control (Algorithm~\ref{alg:admission}) and the sequential optimization technique to solve $\mathcal{P}_0$.
	
	Interestingly, Proposed achieves solution quality very close to the CVX benchmark, with only 2.2\% maximum deviation, despite significant computational complexity difference: $\mathcal{O}(T_{\max} SK)$ compared to $\mathcal{O}(|\mathcal{K}_{\text{adm}}|^3)$~\cite{boyd2004convex}, where $|\mathcal{K}_{\text{adm}}| \ll K$. This close alignment stems from the convergence behavior of the proposed bandwidth allocation scheme (Algorithm~\ref{alg:bandwidth}), which effectively tracks marginal utility gradients and approaches the optimum under $\mathcal{K}_{\text{adm}}$. In contrast, Baseline yields 45\% higher average weighted sum-rate than Proposed at lower UE density because it admits all UEs with uniform allocation (avoiding rejection penalties). However, as UE density increases, round-robin spreads bandwidth evenly while Proposed maintains focused allocation on $\mathcal{K}_{\text{adm}}$, narrowing the gap to 7.8\% at $K = 100$. Notably, although admission control reduces average weighted sum-rate by prioritizing URLLC UEs at the expense of eMBB throughput, this trade-off substantially improves the average success rate, as detailed in Section~\ref{Average Success rate Performance}.
	
	\subsubsection{Average Success Rate Performance}
	\label{Average Success rate Performance}
	
\begin{figure}[tb]	
		\centering
		\includegraphics[scale=0.2]{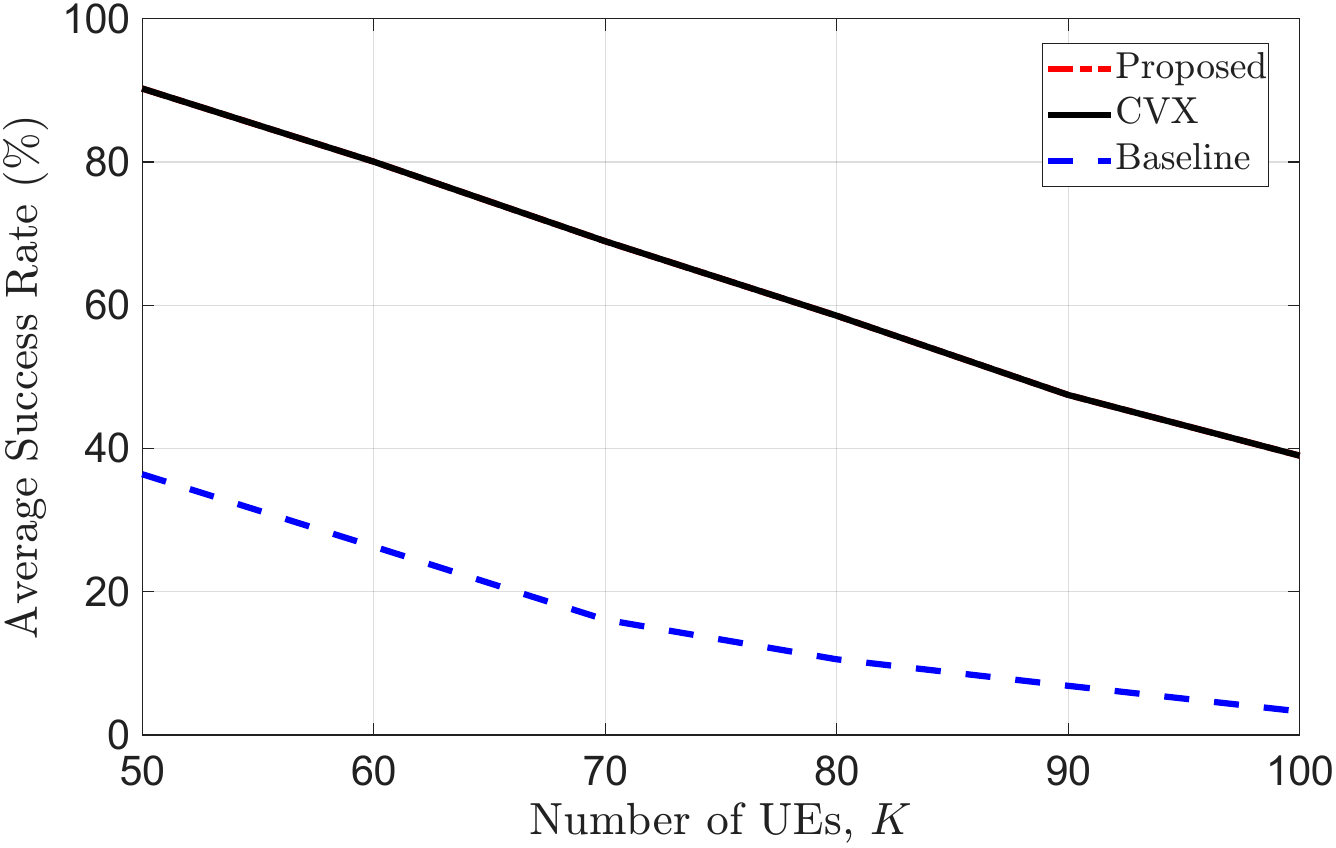}
		\caption{Average success rate obtained by various schemes for eMBB UEs with $ \tau_p = 10 $ and $M = 100$ APs.}
		\label{Success_embb}
	\end{figure}

\begin{figure}[tb]	
		\centering
		\includegraphics[scale=0.2]{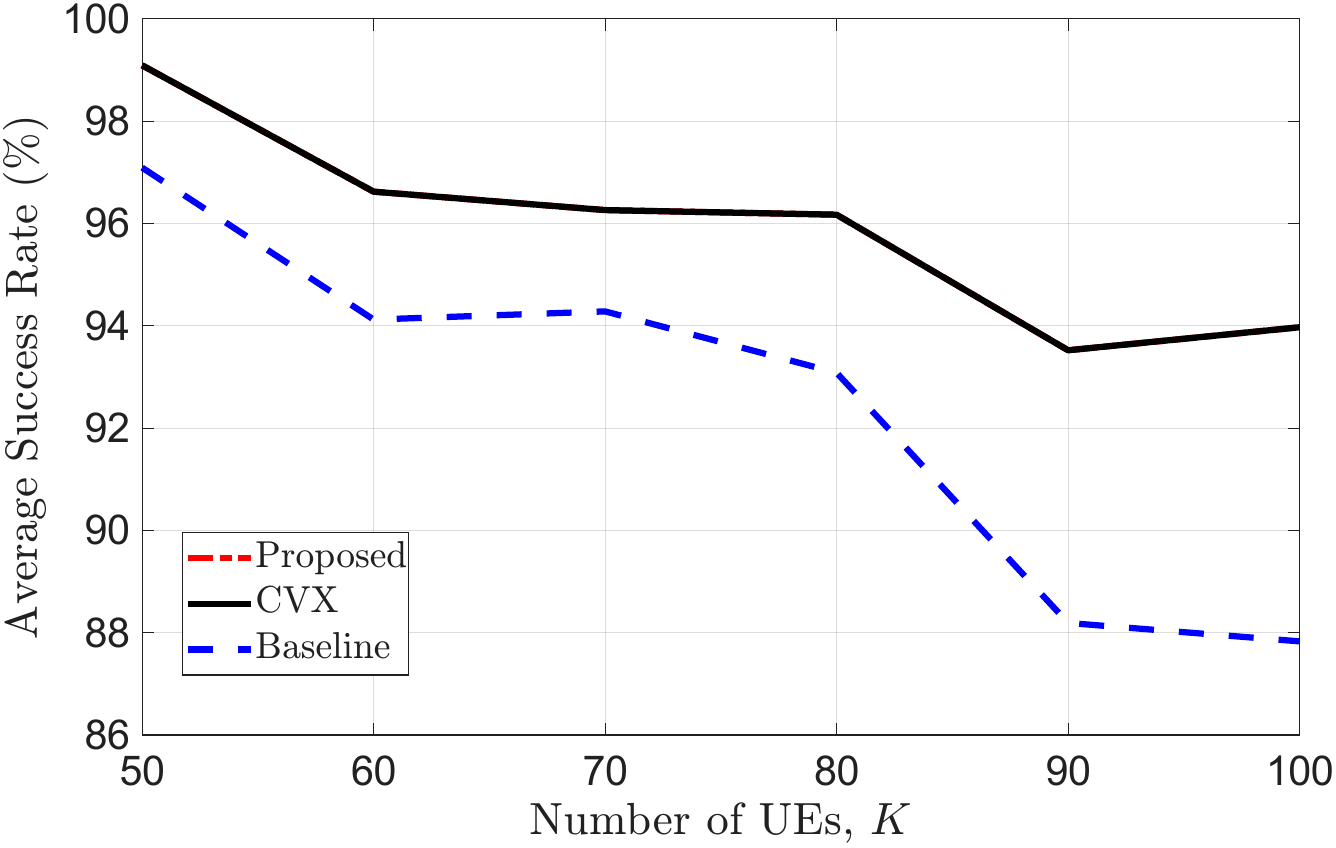}
		\caption{Average success rate obtained by different schemes for URLLC UEs with $ \tau_p = 10 $ and $M = 100$ APs.}
		\label{Success_urllc}
	\end{figure}

	Figs.~\ref{Success_embb} and~\ref{Success_urllc} illustrate the average success rate, defined as the fraction of UEs satisfying their QoS constraints (minimum-rate for eMBB, delay for URLLC). Importantly, Proposed and CVX achieve identical success rates, confirming that Algorithm~\ref{alg:bandwidth} preserves QoS feasibility while matching the solution quality nearly identical to the CVX optimum. Compared to Baseline, Proposed outperforms it by 1085\% for eMBB and 7\% for URLLC. Note that URLLC success rates consistently exceed eMBB due to smaller bandwidth demands per packet and URLLC-prioritized admission control. As $K$ increases, success rates degrade for all schemes due to resource contention, however, Proposed maintains substantially higher QoS compliance. This advantage stems from the synergy between Algorithm~\ref{alg:admission} and Algorithm~\ref{alg:bandwidth}, which jointly allocate bandwidth based on slice-specific weights and SE, directing resources toward high-priority UEs capable of meeting stringent constraints.
	
	\subsubsection{Computational Complexity Performance}
	\label{Computational Complexity Performance}
	
\begin{figure}[tb]	
		\centering
		\includegraphics[scale=0.2]{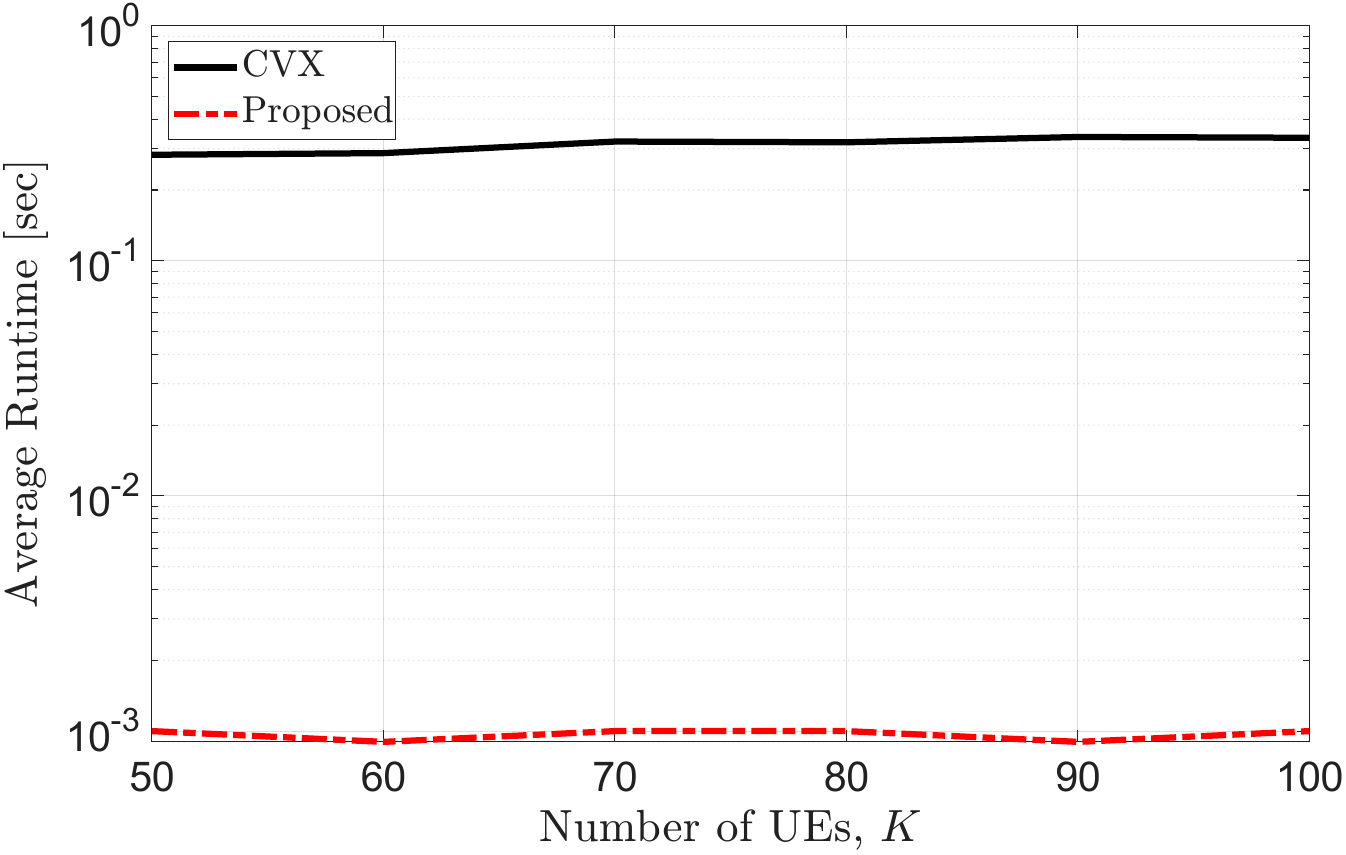}
		\caption{Average runtime comparison between CVX and the Proposed schemes for varying UE $K$ with $M = 100$ APs.}
		\label{runtime}
	\end{figure}

	Fig.~\ref{runtime} reports the average runtime comparison between CVX and the Proposed schemes. As explained in Section~\ref{Average Weighted Sum-rate Performance}, the Proposed approach achieves a substantial reduction in computational complexity, reaching up to two orders of $K$ relative to CVX. In particular, it reduces runtime by up to 99.7\% while still delivering nearly the same weighted sum-rate and maintaining comparable QoS satisfaction performance. Both schemes exhibit stable runtime across all UE densities because admission control (Algorithm~\ref{alg:admission}) filters only feasible candidates before bandwidth allocation, thereby limiting the problem size independent of $K$. This stability, combined with the lower complexity of Algorithm~\ref{alg:bandwidth}, underscores the computational efficiency and practical viability of the Proposed scheme for QoS-critical, resource-constrained large-scale CF mMIMO deployments.	
	
	\subsubsection{Sensitivity Analysis of the Proposed Solution}
	\label{Sensitivity Analysis of the Proposed Solution}
	
\begin{figure}[tb]	
		\centering
		\includegraphics[scale=0.3]{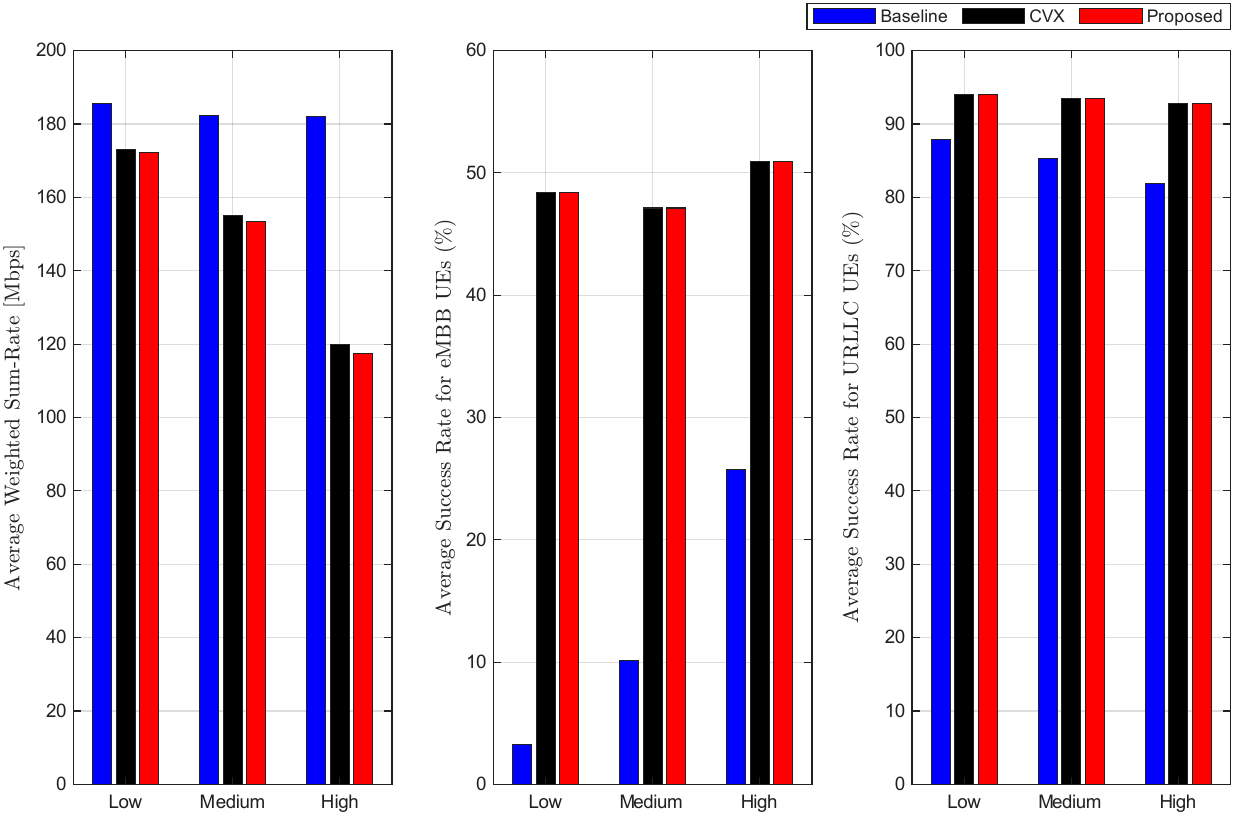}
		\caption{Sensitivity analysis of the different schemes under varying load with $K = 100$ UE and $M = 100$ APs.}
		\label{sensitivity}
	\end{figure}

	Fig.~\ref{sensitivity} examines the sensitivity of the Proposed scheme under varying slice compositions, where Low, Medium, and High loads correspond to 70\%/30\%, 50\%/50\%, and 30\%/70\% eMBB/URLLC UE ratios, respectively. Notably, Proposed maintains near-optimal solution quality, deviating from CVX by at most 2.0\% across all loads, validating the effectiveness of Algorithm~\ref{alg:bandwidth}. Compared to Baseline, the Proposed scheme intentionally sacrifices average weighted sum-rate by 7.3\%, 15.7\%, and 35.5\% at Low, Medium, and High loads, respectively, to guarantee QoS compliance. However, this modest sum-rate penalty yields substantial QoS benefits. For eMBB UEs, Proposed achieves 1371\%, 365\%, and 98\% higher success rates than Baseline at Low, Medium, and High loads, respectively, while maintaining stable performance (47-51\%) across all loads. For URLLC UEs, Proposed outperforms Baseline by 7.0\%, 9.5\%, and 13.3\%, sustaining 93-94\% success rates even under High load with 70\% URLLC demand. This behavior demonstrates that the URLLC-prioritized admission control (Algorithm~\ref{alg:admission}), complemented by Algorithm~\ref{alg:bandwidth}, effectively adapts to diverse slice mixes by directing resources toward high-priority, latency-sensitive UEs, thereby ensuring robust QoS compliance at the expense of raw sum-rate.

	\vspace{-4mm}	
	\section{Conclusion}
	We develop a computationally efficient joint optimization framework that combines hierarchical admission control and gradient-based bandwidth allocation to address the NP-hard problem of satisfying heterogeneous QoS constraints across eMBB and URLLC slices in NS-enabled UC CF mMIMO systems under bandwidth scarcity where aggregate demand may exceed available capacity. By selectively admitting UEs and prioritizing URLLC through admission decisions while reallocating bandwidth via marginal utility gradients, the proposed framework intentionally trades sum-rate for QoS compliance, achieving near-CVX solution quality with significantly reduced complexity. Numerical results prove that the proposed heuristics are robust across varying eMBB/URLLC traffic compositions, maintaining consistent success rates even under high URLLC demand. These results further confirm the viability of computationally efficient, priority-aware resource management for latency-critical applications in bandwidth-constrained large-scale CF deployments. Incorporating joint power allocation to enhance SE and QoS resilience under stringent resource constraints remains a promising direction for future work.
	
	\vspace{-3mm}
	\bibliography{Jan31_2022}
	\bibliographystyle{ieeetr}
	
\end{document}